# Data Driven Reference Architecture for Smart City Ecosystems

Mohammad Abu-Matar
EBTIC
Khalifa University of Science & Technology
Abu Dhabi, UAE
Mohammad.abu-matar@kustar.ac.ae

John Davies
BT Research & Innovation
British Telecommunications
Ipswich, UK
John.nj.davies@bt.com

*Abstract*— With the convergence of information and telecommunication technologies, the vision of the 'Smart City' is fast becoming a reality. City governments in a growing number of countries are capitalizing on these advances to enhance the lives of their citizens and to increase efficiency and sustainability. In this paper, we elaborate on smartCityRA, a reference architecture for Smart City projects, which serves as the design language for creating smart cities blueprints. Such a blueprint caters for diverse stakeholders, devices, platforms, and technologies. We report on our experience in carrying out a proof-of-concept use case with a major telecommunication provider in the UAE. In doing so, we refined our multiple-view model of the initial smartCityRA reference architecture. We show that Data in smart city applications drive the entire development lifecycle and should be considered early in the development cycle. In addition, Data affects all the other views in the smartCityRA and hence the Data View needs to be at the heart of the entire smartCityRA. Realizing the Data view using a component like a Data Hub helped in creating a central integration location for disparate data from different sources, thus reliving developers from dealing with several entities individually. Finally, we show that any smart city reference architecture, like smartCityRA, should be at the right level of abstraction to enable the flexibility of adoption and adaptation by different stakeholders and components.

## I. INTRODUCTION

Many city governments [1], [2] are increasingly embracing the concept of smart cities in order to increase operational effectiveness and to satisfy the needs of their citizens and businesses. A smart city is defined as a city ecosystem that leverages connected technologies for the purpose of the betterment of city operations and intelligent near real-time decision making [3]. Gartner, Inc., a research and advisory firm, defines a smart city as "an urbanized area where multiple sectors cooperate to achieve sustainable outcomes through the analysis of contextual, real-time information shared among sector-specific information and operational technology systems." [4] Clearly, such an ecosystem comprises of several heterogeneous and highly distributed cyber-physical [5] subsystems where diverse stakeholders need to interact and collaborate.

As we have planned in our smart city reference architecture (smartCityRA) paper [6], we conducted a use case utilizing BT's Data Hub [7], [13] as the main data integration platform. Our aim was to discover what activities application developers had to go through in order to create applications in the wider smart city ecosystem. In doing so, we have also refined the smartCityRA multiple-view meta-model that we created in the initial paper [6]. Specifically, we describe the refinements of the Data View of the meta-model based on the use of the BT Data Hub [7] in addition to suggestions on other views.

We explain the refinement process by describing a use case we performed for United Arab Emirates (UAE) weather measurements in collaborations with the main telecommunication provider in the country.

We show that data in smart city applications drive the entire SDLC (software development lifecycle) and should be considered early in the development cycle. In addition, data affects all the other views in the smartCityRA and hence the Data View would serve as the heart of the entire smartCityRA. Realizing the Data View using a component like a Data Hub [7] helped in creating a central integration location for disparate data from different sources, thus reliving developers from dealing with several entities individually. Finally, we show that any smart city reference architecture, like smartCityRA, should be at the right level of abstraction to enable the flexibility of adoption and adaptation by different stakeholders and components.

## II. UAE WEATHER USE CASE FOR A TELCO PROVIDER

The main telecommunication provider in the UAE engaged us to create a proof-of-concept trial for IoT and smart city applications using the BT Data Hub. The telco provider has purchased a leading IoT software platform and they needed help in developing their IoT service and application delivery process. The telco provider stakeholders consisted of several groups with varying concerns that spanned software, hardware, networking, sensor device management, and policy makers. This multiple stakeholders landscape was an interesting example of the typical ecosystem nature of smart city applications and a testament to the need of a systematic approach and strategy for delivering smart city offerings, providing an excellent test case for our reference architecture.

We settled on a relatively simple case study of UAE's weather data. In essence, we wanted to aggregate weather data in the UAE into one repository in order to provide the Telco provider access to this data easily without having to deal with several weather sources in each emirate. Once populated, this data could be used by developers to create their own applications using the provided API without worrying about the low level sensor types, protocols and data formats. We will use this use case as a running example throughout the paper.

### III. SMARTCITYAR META-MODEL

We briefly describe the initial smartCityRA meta-model [6] which we will use as a basis for the use case in this paper. To formalize the process of creating smart city applications and services, we created an initial multiple-view meta-model based on ISO/IEC/IEEE 42010 standard [8] for reference architectures and Multiple-View Modeling [9]. SmartCityRA comprises relevant software-defined views and modeling elements. These views were a first attempt (non-exhaustive) in the iterative building of the smartCityRA.

The meta-model identifies nine views, their corresponding elements, and the relationship among the views. All views are unified by a Capability View which represents the business requirements provided by a smart city project. This unification is similar to the seminal work of Kruchten [9] where he introduced the '4+1' view model of software architectures, in which he advocated a multiple view modeling approach in which the use case view is the unifying view (the 1 view of the 4+1 views). Thus, the Capability View in our smartCityRA meta-model is the unifying view that refers to Weather Data in our use case. We provide a brief description of each view as depicted in Fig. 1:

1) Capability View – this is an abstract representation of any capability, i.e. feature, provided by a smart city system. It is abstract in the sense that it could be specialized to represent a capability of the whole city, of a subsystem, a department, an application, or even city-wide goals.

2) Participation View – this view models several types of stakeholders: Citizens, Policy Makers, Businesses, etc. Like all other views in the meta-model, this is an abstract view that could be specialized to model any type of participants.

3) Place View – A Place models buildings, hospitals, municipalities, and the like.

4) Service View – this view models specific services provided by smart city applications. A Service has a contract and an interface. The Service Interface is provided by the Service Provider (Participant in the Participation View), and the Service Contract is used by both the Provider and the Consumer of the Service.

5) Data View – This view, the focus of this paper, models any type of data involved in smart city applications. Exchange of data is at the heart of the smart city ecosystem and this view could be specialized to model many data types. We explain in this paper how this view was realized by the aforementioned Data Hub.

6) Application View – this view could model: individual applications, system components, enterprise systems, or any computation unit that provides the functionality needed by services.

7) Infrastructure View – this view models physical components that underlie the smart city ecosystem. It has an abstract Device class that could be specialized into any type of device (Fig. 1). This view is not intended to be complete, rather a placeholder for any infrastructure related elements, which could include Movement Sensor, Actuators, Communication Links, and the like.

8) Business Process View – this view models the business flow of interdependent business tasks executed on behalf of business, organizations, and the like. Business Processes could be realized by Services, Applications, and Individuals as depicted in the meta-model (Fig. 1). A smart city ecosystem will have many business processes executed across many boundaries and realized by disparate applications and services.

9) Analytics View – this view models the types of analytics required to achieve the goals of smart decisions. It should be noted that this view was added in this research based on our experience in the UAE use case as explained below.

Table I lists the smartCityRA views along with their Stakeholders, Concerns, and Model Type. Fig. 1 depicts a high level UML meta-model for the views, their corresponding elements, and their relationships.

### IV. DATA DRIVEN SMART CITY META-MODEL

In this section, we detail the meta-model of the Data Hub and show how it fits within the overall smart city reference architecture. We show that by using and understanding the components of the Data Hub, via the use case, we were able to refine and further articulate what should be contained in the Data View of smartCityRA.

#### A. BT Data Hub

The BT IoT data hub [7], [13] enables information from a very wide range of IoT and other sources to be brought onto a common platform and presented to users and developers in a consistent way. Its portal provides an interface through which users can browse an IoT data catalogue and select and subscribe to data feeds that they want to use. Aggregating data in this way offers economies of scale and lowers the barrier to participation in an IoT ecosystem, thus fostering innovation, as well as breaking down data silos.

An API enables access to feeds, secured by APIs keys, from browsers or from software. A relational, GIS capable, database enables feeds to be filtered according to a wide range of criteria. The edge adapter layer enables information coming onto the Data Hub to be converted to a standard format. It also provides a consistent API to users (typically application developers). The data held on any given instance of the hub is catalogued in a machine-readable Hypercat catalogue [10], [14].

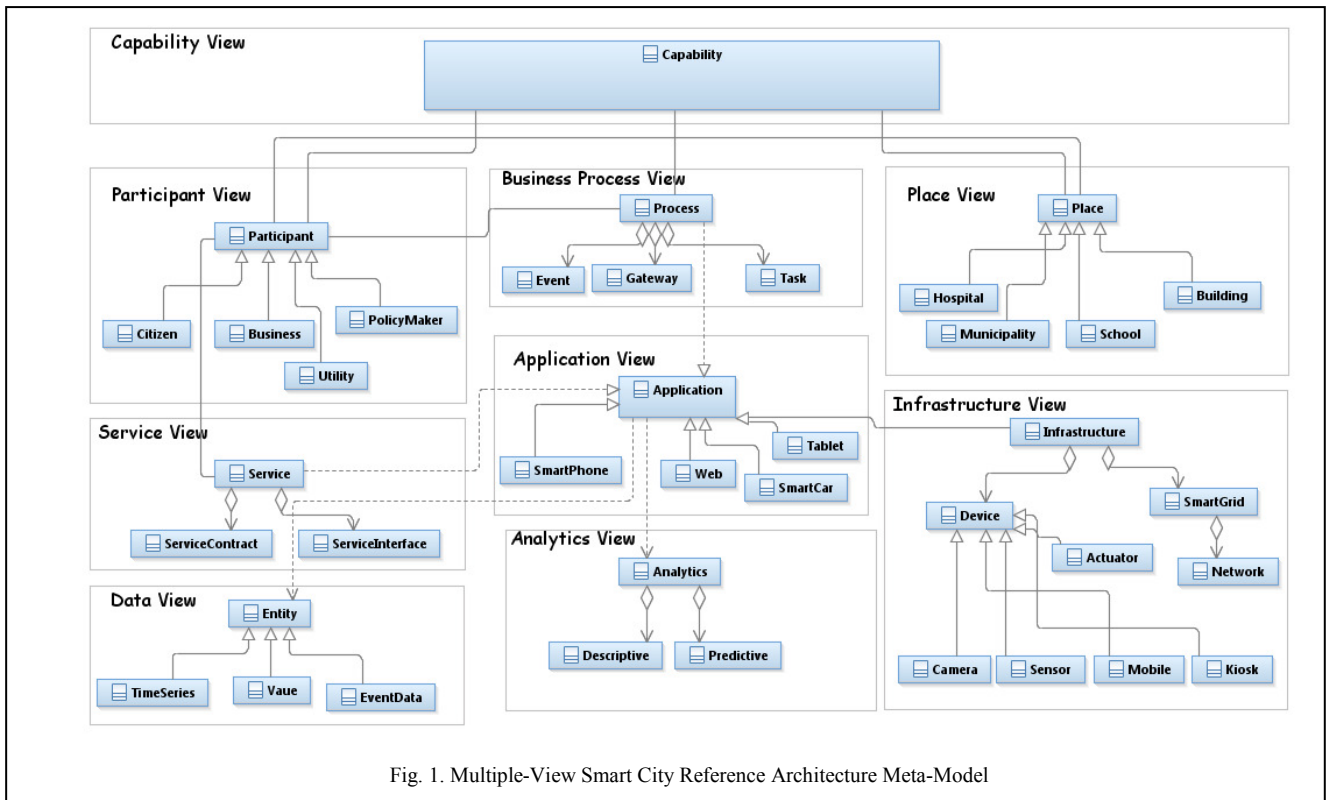

Fig. 1. Multiple-View Smart City Reference Architecture Meta-Model

Data Providers are able to easily publish data feeds and advertise availability of their data in some form of catalogue that allows potential users to discover and use their data. They can control access to their data and provide any guarantees they feel are appropriate on data availability and quality.

The platform enables end users and application developers to discover which applications are available based on their particular interests and requirements, geographical or topical both current and historical. The high level design of the BT Data Hub is shown in Fig. 2.

*B. BT Data Hub Meta-Model*

In the initial smartCityRA meta-model [6], we started with an abstract Data View knowing that this view will need to be refined. By using the Data Hub as our main data integration repository, we learned a great deal about what a Data View should contain. Hence, we created a meta-model for the Data Hub to extract the main elements that should be present in the Data View. Fig. 3 below depicts the main elements of the hub meta-model.

At the heart of the data hub meta-model is the Information element. Information element is specialized into Measurement (e.g. time series sensor data), Event (e.g. announcement, traffic accident), and Context (e.g. geolocation) elements.

Information element is exposed via the API element, which is specialized into Provider and Developers elements. In our case, these APIs are realized by REST HTTP protocol, however other implementation may use other access protocols and still conform to the meta-model.

The Information and API elements are advertised by the DataDiscovery element, which is specialized into the Portal and Catalog elements. Information in the data hub can be discovered manually using the Portal, or programmatically using the Catalog APIs.

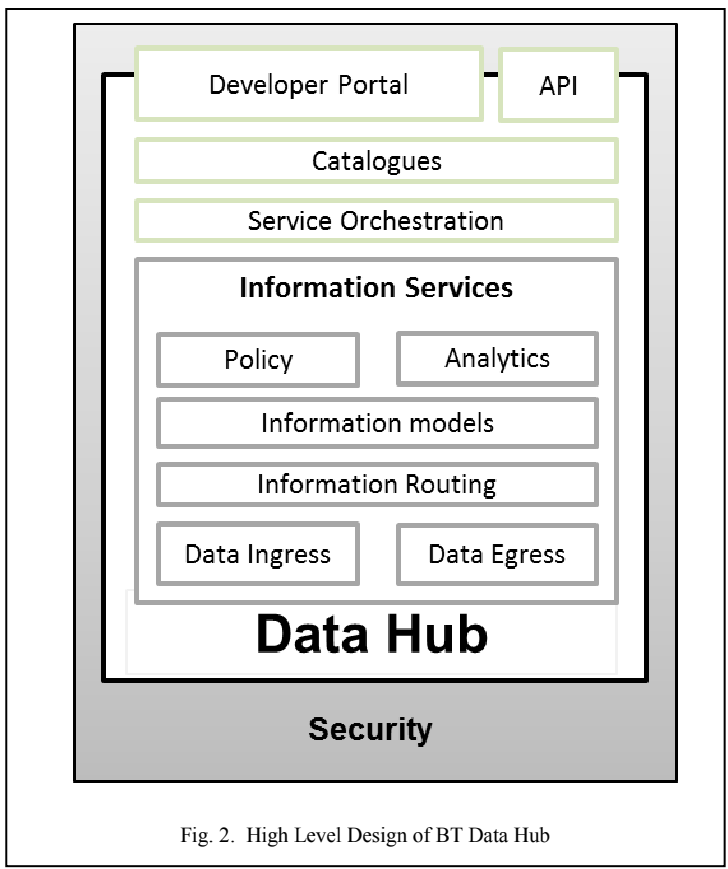

Fig. 2. High Level Design of BT Data Hub

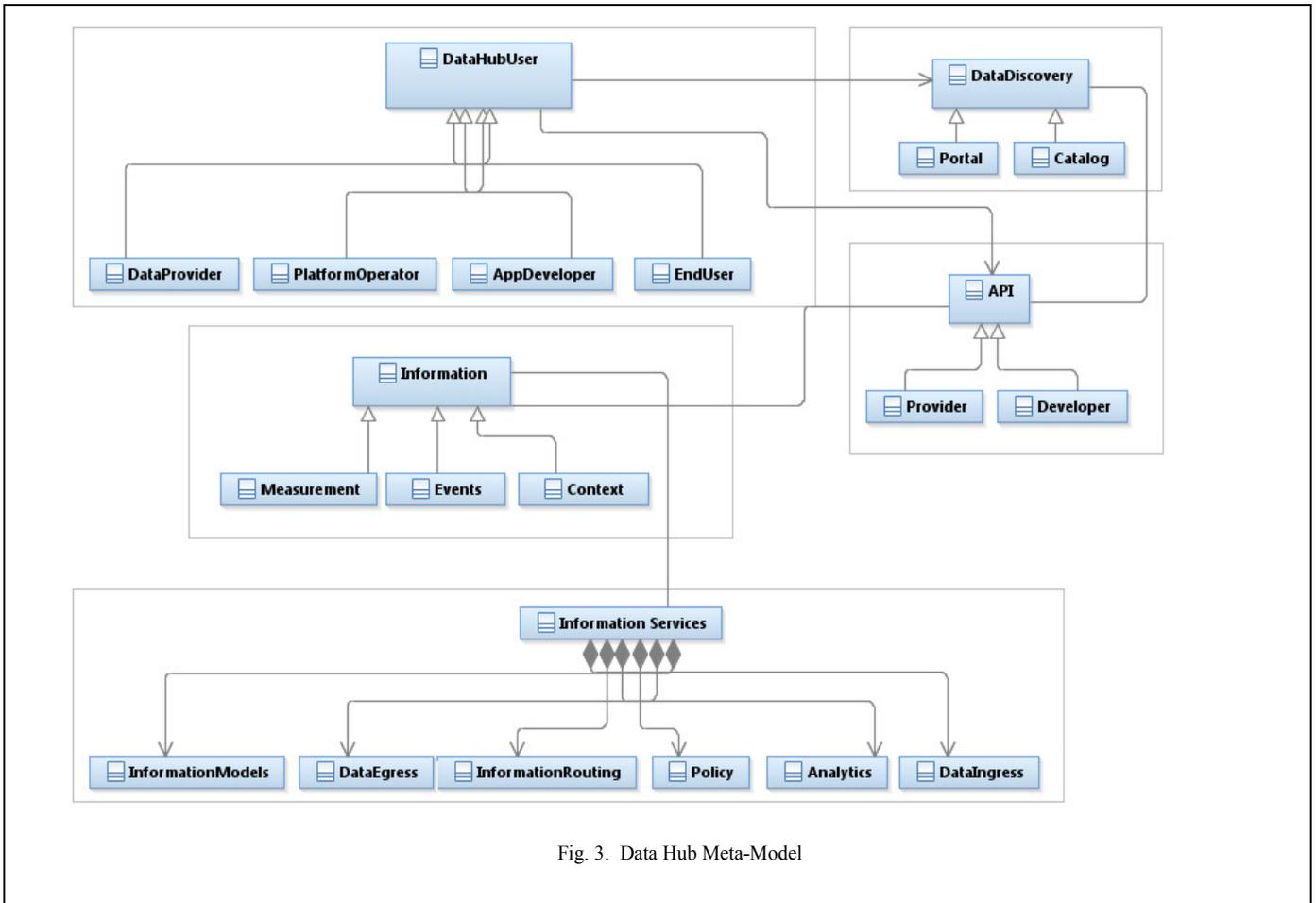

Fig. 3. Data Hub Meta-Model

DataHubUser element describes different roles that interact with the data hub. Namely, DataProvider, PlatformOperator, AppDeveloper, and EndUser elements.

Finally, the inner-workings if the data hub are modeled by the InformationServices element, which is specialized into InformationModel, DataEgress, DataIngress, InformationRouting, Policy, and Analytics elements. The description of the internal components of the hub is out of scope of this paper.

## V. SDLC FOR THE UAE WEATHER USE CASE

In this section we describe the steps taken to prepare and use the weather measurements data in the weather use case. As we describe each step, we explain what views from the smartCityRA were exercised. Our aim is to show that smart city/IoT application development involves multiple aspects that are interrelated, i.e. views. In addition, we hope to show that developers can benefit from starting with an initial blueprint like the smartCityRA.

### A. Preparing Measurements Data

We obtained CSV formatted real-time weather data from the Weather Underground website [11] for all emirates in the UAE. We used URLs with specific parameters to get the weather data from different stations on an hourly basis.

Once the required data was identified (in this case, weather data), we needed to create the relevant feeds in the Data Hub. We created a sensor feed for each emirate in the UAE. Then, we created a data stream for each weather measurement per feed, e.g. temperature, humidity, precipitation, etc. Fig. 4 depicts the data streams for the Abu Dhabi weather data feed. In this step, the following views from the smartCityRA are exercised:

- Data View – where the data hub constitutes the entire view.
- Participant View – where Platform Operator was responsible for creating the feeds.

### B. Populating Data into Feeds

Since the Data Hub accepts measurements data in EEML format [12], we had to write an edge adapter that converts CSV to EEML. Then, we used the Data Hub's Providers REST APIs to feed the data into the hub. As in the previous step, the smartCityRA Data and Participation Views were exercised in this step.

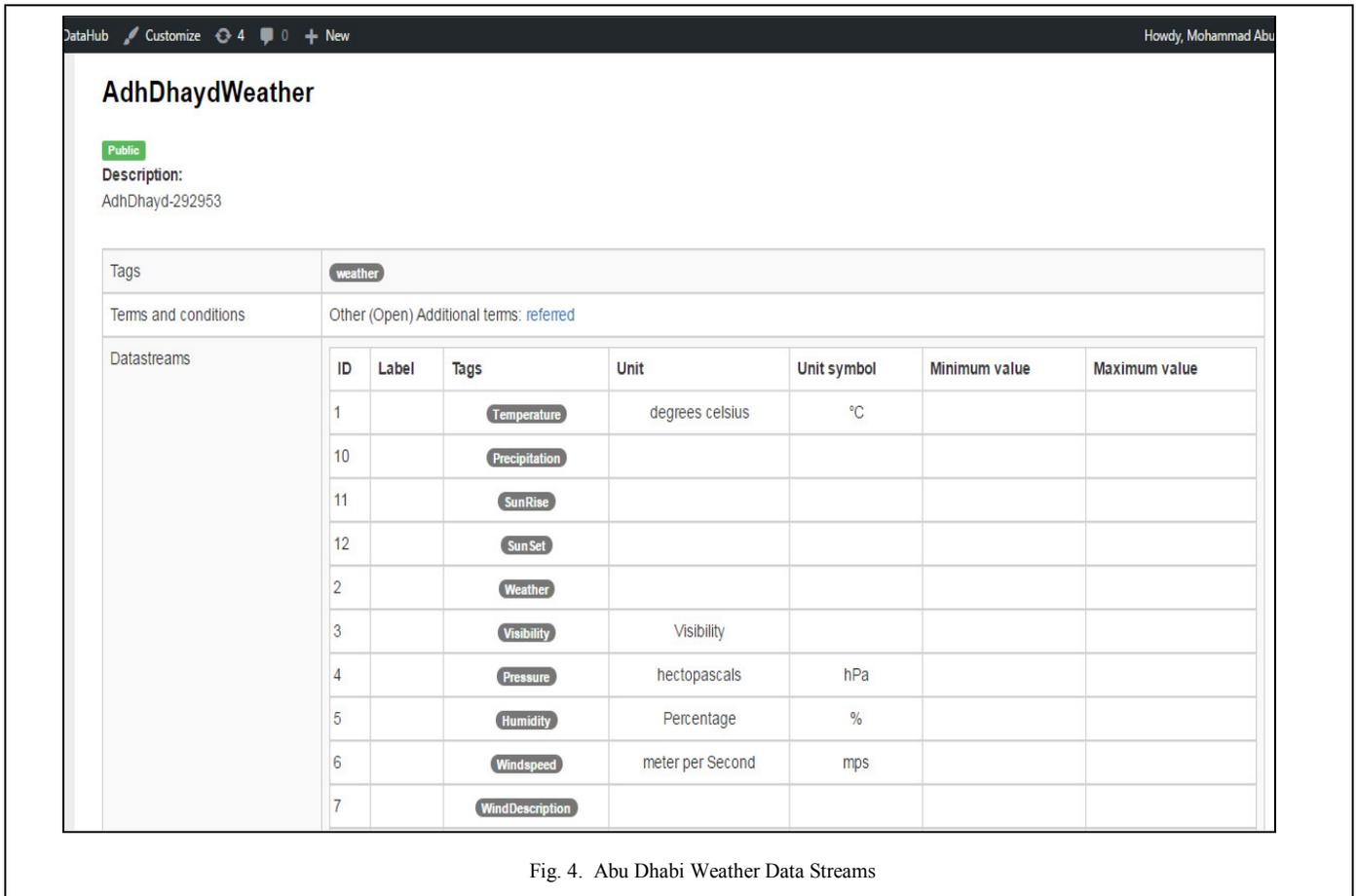

Fig. 4. Abu Dhabi Weather Data Streams

*C. Application Development*

Once measurement data is populated and configured, developers can browse the data hub catalog, subscribe to feeds, and use the REST APIs to get sensor (and other) data. In addition to the Data View, the following views from the smartCityRA are exercised:

- Application View – where developers design and implement their application by using the data hub REST APIs. Here, different developers can use the same APIs to create different applications.
- Service View – this view represent the REST APIs for the data hub.
- Business Process View – this view is exercised if the application triggers or uses a business process workflow. For example, if the weather is fogy in a certain day, this might triggers road closure throughout the city and hence a specific process has to be followed.
- Participant View – where app designers, developers, and testers are participating in the overall app development.
- Place View – in this use case, the Place View represent the entire UAE, the different emirates, and the individual cities.

It should be noted that the Infrastructure View is implicitly used in the app development lifecycle, but not exercised directly in this use case since sensor data is abstracted from the physical layer.

VI. REFINED SMART CITY REFERENCE ARCHITECTURE

From the outset, we intended the devotement of the smartCityRA to be iterative. Hence, we designed the reference architecture to be abstract and extensible so it can be refined by the development community.

By utilizing the data hub in the weather use case, we refined the Data View in the initial smartCityRA meta-model. We learned that we should adopt a data-first approach, since getting and preparing the data involve several stages and challenges. Hence, the refined Data View consists of the following main elements (Fig. 3):

- Data Ingress – describes how data should be inputted into the hub and in what format. Data is fed into the hub via a REST APIs.
- Data Egress – describes how data is outputted from the hub and in what format. Data is read from the hub via REST APIs.
- Information is modeled as Feeds, which can be Measurements (sensors), Events, and Contextual Data.

- Data Discovery – data is advertised through a portal and could be discovered programmatically by REST APIs.
- API element is mapped to the Service View in the initial smartCityRA (Fig. 1).
- DataHubUser element and its children are mapped to the Participant View.

Based on our conversations with several stakeholders during the use case, Analytics always materialized as a separate perspective from the Application View. In essence, the same application can use different analytics engines and techniques. Hence, we added a new Analytics View that is connected to the Application View in the smartCityRA meta-model (Fig. 1), (Table I).

Security concerns, as expected, came up in several discussions during the use case. Since this was a proof-of-concept prototype, we did not elaborate on security issues beyond the obvious authentication and authorization issues. However, based on the usage of the data hub and sensor data management, we gleaned the following insights to be used in future refinement of smartCityRA:

- Data hub has the expected authentication and authorization capabilities for the hub users: administrators, operators, providers, and developers.
- Once sensor feeds are created, as described above, the hub automatically creates a security key that needs to be submitted with every REST API call. Keys are shared with authorized developers.
- Low level physical sensor security is assumed and should be handled by infrastructure teams. This assurance is important to have at the beginning of the SDLC since software developers are shielded from the low level sensor details by the hub.
- Security spans all views within the smartCityRA meta-model; hence a Security View should be added to the meta-model in future iterations.

A. Lessons Learned

We summarize some lessons learned during the use case:

- The telco provider's stakeholders consisted of several teams with varying background that included both technical and strategic teams. As a result, communicating and sharing ideas proved to be challenging at the beginning. We realized that starting with a blueprint, i.e. reference architecture, could facilitate the discussions and reap faster benefits.
- The telco provider has already purchased a leading IoT platform before starting the use case. A great deal of the discussion has centered on the comparison between the data hub and their platform. The details of the comparison is out of the scope of this paper, but using the smartCityRA proved to be valuable because it enabled us to explain the roles of the IoT platform and the hub logically and independent form technology. Once, these components were placed within the appropriate views, it was easy to use them

TABLE I. SMART CITY REFERENCE ARCHITECTURE VIEWS

| View | Stakeholders | Concerns | Model Type |
|---|---|---|---|
| **Capability** (Unifying View) | Citizens, Policy Makers, Business | Citizens Satisfaction, Compliance | Use Case Diagram |
| **Service** | Citizens, Policy Makers, Auditors | Realize Capabilities | Service Architecture Diagram |
| **Participation** | Citizens, Policy Makers, Service Providers | Ease of Citizens Interaction, Quality Provided Services | Interaction Diagram |
| **Data** | Devices, Developers, IT | Data Quality, Data Accuracy, Data Integrity, Timeliness | Class, Data Flow diagrams |
| **Analytics** | Business Analysts, Policy Makers | Decision Making, Business Insights, Actions | Mathematical, Algorithmic, Procedural |
| **Application** | Developers, IT | Business solutions | Component and Class diagrams |
| **Infrastructure** | Utilities | Availability, Uptime | Deployment diagram |
| **Business Process** | Business Analysts | Business goals | Activity diagram |
| **Place** | Citizens | Location of Participants. | Class diagram |

- both in the same use case.
- Data in smart city applications drive the entire SDLC and should be considered early in the development cycle. Data affects all the other views in the smartCityRA and hence the Data View is at the heart of the entire smartCityRA.
- Realizing the Data view in smartCityRA by using a component like the Data Hub helped in creating an integration central location for disparate data from different sources, thus reliving developers from dealing with several entities individually.
- Any smart city reference architecture, like smartCityRA, should be at the right level of abstraction to enable sufficient flexibility for adoption and adaptation by different stakeholders and components.

## VII. CONCLUSION AND FUTURE WORK

As smart city initiatives gather momentum worldwide, it is increasingly evident that a new approach for designing such ultra large and ultra-heterogeneous ecosystems is needed. In previous research [6], we proposed a reference architecture, smartCityRA, inspired by several disciplines in software engineering such as SOA. SmartCityRA serves as a blueprint and starting point that contains architectural building blocks, best practices, and patterns; instead of starting from scratch.

As we anticipated in our smartCityRA paper [6], we conducted a use case utilizing BT's Data Hub [7], [13] as the main data integration platform. In doing so, we learned what activities application developers had to go through in order to create applications in the wider smart city ecosystem. Additionally, we refined the smartCityRA multiple-view meta-model that we created in the initial paper [6]. Specifically, we described the refinements of the Data View of the meta-model based on the use of the Data Hub in addition to suggestions on other views.

We showed that Data in smart city applications drive the entire SDLC and should be considered early in the development cycle. In addition, Data affects all the other views in the smartCityRA and hence the Data View is at the heart of the entire smartCityRA. Realizing the Data view using a component like a Data Hub helped in creating a central integration location for disparate data from different sources, thus reliving developers from dealing with several entities individually.

For future work, we intend to conduct more use cases with local stakeholders in the UAE to further refine smartCityRA. In addition, we are in the early planning stages to create a reference implementation model for smartCityRA. Such a model would serve as a Demonstrator implementation for other developers to rely on if they follow smartCityRA. This effort could lead to the creation of an Integrated Development Environment (IDE) that embodies the smartCityRA framework with simulation capabilities to test early design scenarios.

Finally, recent trends in model driven engineering (MDE) [15] advocate the construction of Domain Specific Languages (DSL) instead of relying on general purpose modeling languages, e.g. UML, to attain productivity, ease of communication, and expedite development. We envision the development of a smart city modeling language (smartCityML) based on the smartCityRA meta-model presented in this paper. This language could have several benefits: 1) Ease of use for diverse specializations of stakeholders; 2) Code generation opportunities for standardized sensors and devices; 3) Quick prototyping to cater for policy makers; and 4) Simulation of design prior to construction. ThingML [15] DSL is a promising language that aims to abstract connected devices in any IoT application. smartCityML could use ThingML for the Infrastructure View defined in the smartCityRA meta-model above.